%% file: main.tex
\documentclass[sigconf]{acmart}
\pdfoutput=1

\usepackage{booktabs} 
\usepackage{graphicx}
\usepackage{subcaption}
\usepackage{xcolor,soul}

\acmArticle{4}
\acmPrice{15.00}

\begin{document}
\title{Exploiting Population Activity Dynamics to Predict Urban Epidemiological Incidence}

\author{Gergana Todorova}
\authornote{Gergana Todorova was placed as a data scientist at the North West Ambulance Service, NHS during the period of the present work.}
\orcid{1234-5678-9012}
\affiliation{%
  \institution{Lancaster University}
  \city{Lancaster}
  \state{United Kingdom}
}
\email{gdtdrv@gmail.com}

\author{Anastasios Noulas}
\affiliation{%
  \institution{Center for Data Science, New York University}
  \city{New York}
  \state{USA}
}
\email{noulas@nyu.edu}

\begin{abstract}
Ambulance services worldwide are of vital importance to population health. Timely responding to incidents by dispatching an ambulance vehicle to the location a call came from can offer significant benefits to patient care across a number of medical conditions. Moreover, identifying the reasons that drive ambulance activity at an area not only can improve the operational capacity of emergency services, but can lead to better policy design in healthcare. In this work, we analyse the temporal dynamics of 5.6 million ambulance calls across a region of 7 million residents in the UK. We identify characteristic temporal patterns featuring diurnal and weekly cycles in ambulance call activity. These patterns are stable over time and 
across geographies. Using a dataset sourced from location intelligence platform Foursquare, we establish a link between the spatio-temporal dynamics of mobile users engaging with urban activities locally and emergency incidents. We use this information to build a novel metric that assesses the health risk of a geographic area in terms of its propensity to yield ambulance calls. Formulating then an online classification task where the goal becomes to identify which regions will need an ambulance at a given time, we demonstrate how semantic information about real world places crowdsourced through online platforms, can become a useful source of information in understanding and predicting regional epidemiological trends.
\end{abstract}

\maketitle

\input{intro}

\input{related}
\input{prelim}
\input{analysis}
\input{urbanrisk}
\input{prediction}

\input{conclusion}

\bibliographystyle{ACM-Reference-Format}
\bibliography{biblio} 

\end{document}

%% file: intro.tex
\section{Introduction}
Changes in lifestyle as well as intensifying urbanisation and travel not only pose an increased health risk, but add extra pressure to emergency services that operate day and night in the world's largest cities. In the United Kingdom, the number of health related incidents that require urgent response has been steadily increasing, resulting to an unprecedented number of calls that require an ambulance response~\cite{nhsreport}. As a result, during 2015 only 5 out of 11 ambulance trusts operating in England have met their operational target to respond to $75\%$ of most urgent calls within eight minutes~\footnote{\url{https://www.bbc.com/news/health-33166204}}. Incidents such as \textit{cardiac arrest} require immediate attention by paramedic specialists, with ambulance response times  being critical to patient survival rates~\cite{pell2001effect}. Consequently, intelligence on optimally deploying paramedic crews across geographic areas and also with appropriate timetabling becomes a significant challenge for ambulance services worldwide. 

To meet the goal of timely responding to urgent health incidents ambulance organisations over the past decade have been moving towards integrating computational methods and data science practices in their planning operations~\cite{henderson2005ambulance}. Such practices include  the recruitment of analysts and data scientists, the development of data warehousing infrastructure, as well as the use of software and cartography tools that support planning allowing for the identification of optimal paramedic crew scheduling~\cite{mason2013simulation}. However the functionality of the class of optimisation tools that simulate emergency event activity is often opaque. They are being frequently packaged in proprietry software becoming effectively a black box for end users~\footnote{\url{https://www.intermedix.com/solutions/response-planning-0}}. In the meantime, scholarly work in environmental epidemiology has been focusing more on accurate statistical modelling of ambulance calls, rather than providing plausible interpretations on the socio-economic and urban activity factors that may drive those. Models developed in this setting 
have used historic frequency information about ambulance calls for instance~\cite{diggle2013statistical, zhou2015predicting} ignoring socio-economic or real time population activity as well as mobility indicators that could also determine the risk for emergency incidents. 

In this work we seek to establish a stronger link between the prevalence of health incidents and spatio-temporal population activities, aiming to bridge the aforementioned gap in the literature. Critically for its novelty, we model activities of local populations using the digital traces emitted by mobile users that engage with location intelligence services like Foursquare. We demonstrate how mobile web indicators of geographic user activity become a useful proxy to understanding the times of the day and days of the week certain urban activities become prominent across city neighborhoods. Moreover, we show how those align across space and time with certain types of ambulance incidents. Finally, 
population activity traces from the web are combined with other data sources such as regional deprivation levels, residence and daytime population volumes as well as historic ambulance calls data, to build an interpretable spatio-temporal prediction framework for emergency incidents. In detail the contributions of the present work are organised as follows:
\\
\textbf{Temporal dynamics of emergency incidents}: Working on a longitudinal dataset of $5.6$ million ambulance calls over a five year period, in Section~\ref{sec:analysis}, we characterise the temporal patterns of various emergency incidents. We observe characteristic temporal signatures on the frequencies of calls for these incidents, with cases such as Chest Pain, Falls or Sick Person peaking during morning hours, when Overdose/Poisoning or Psychiatric and Suicide incidents become progressively more common during evening hours. The diurnal and weekly temporal patterns of natures however remain remarkably stable over time. Moreover, these patterns are similar across geographies with incidents of the same nature to vary over time in a similar manner across different urban neighborhoods or cities.
\\
\textbf{Measuring urban activity risk for emergencies}: In Section~\ref{sec:urbanrisk}, starting from the premise that population level emergencies are driven by the activities people engage with geographically, we study the spatio-temporal relationship between urban activities and ambulance calls. Using check-in information about the visits of mobile users at places collected from the online location intelligence platform Foursquare, we demonstrate that the temporal popularity of certain types of urban activities, going for instance to a nightclub or visiting a water theme park, is strongly associated to an elevated risk for certain types of emergencies such as alcohol poisoning or drowning. 
Based on this evidence and leveraging also on the spatial interactions between ambulance calls and foursquare places, we devise a new metric, named \textit{urban activity risk}, that aims to capture the risk of a geographic region in terms of yielding emergency incidents. 
\\
\textbf{Spatio-temporal prediction of ambulance calls}: 
Finally, in Section~\ref{sec:prediction}, we formulate an online binary classification task where our goal is to predict whether an 
ambulance call will take place at a given area during a specific hour. Despite the inherent sparsity of call events when considering small geographic areas of approximately 1.5 thousand residents, we show how using a combination of information sources about the population characteristics of a given region and historic ambulance activity, predictions that can provide essential intelligence to the ambulance service are feasible. The \textit{urban activity risk} metric we introduce offers useful information in the prediction task outperforming features that are built using socio-economic and demographic data obtained through government surveys. Features mined using online information sources offer an edge in prediction performance pointing to the direction that integrating them in emergency services as proxies to large scale population activity could offer benefits. We further observe how the predictability of ambulance calls varies at different hours of the week with the prediction task being inherently harder during some hours over others, and discuss the implications of this variation in operational and resource allocation terms for the ambulance service. 

Next in Section~\ref{sec:related} we discuss related work on environmental epidemiology, ambulance activity modelling and online health analytics and in Section~\ref{sec:scope} we provide descriptions on the datasets we employ. We conclude in Section~\ref{sec:conclusion}.

%% file: related.tex
\section{Related Work}
\label{sec:related}
\textbf{Mobile Systems and Social Media in Health:} There are two primary research strands that have emerged in recent years in the literature in digital health and are related to our work. Firstly, research using digital traces of users engaging in online social media to solve problems in epidemiology and health analytics, and secondly, the use of smartphone sensing technologies to better track user health status and behaviour and suggest means to improve well being. 

An advantage of using social media to capture population wide epidemiological trends such as obesity rates and eating habits~\cite{abbar2015you, mejova2015foodporn, de2016characterizing} is data descriptions of user activity in real time. At the same time geographic coverage is broad due to the internet scale of online services~\cite{paul2016social, mejova2018online}. Mobile phones complement online social media platforms since they provide a vertical view to user behaviour as mobile application functionality is focused on individuals. Ubiquitous sensing technologies have therefore been exploited to gain deeper views on users' psychological states and develop new frameworks to incorporate this new type of data and tools in clinical psychology practice~\cite{mehrotra2018using, servia2017mobile, canzian2015trajectories, mohr2017personal}. Our work touches the latter class of works as we make use of data crowdsourced through smartphones (Foursquare), but similar to the use of online platforms for health analytics, we work on understanding aggregate population patterns across geographies and times. Our goal here is to use such data as a means to obtain indicators on population activity that can complement existing datasets in healthcare such as the ambulance calls data we analyse in the following sections. To our knowledge this is one of the first works that combines these types of data, building on the line of recent research works that exploit mobility traces of user collectives to understand crime patterns or deprivation levels of local communities~\cite{kadar2018mining,hristova2016measuring, venerandi2015measuring, quercia2014mining}. 
\\
\textbf{Health Geography and environmental epidemiology:} Our work also relates to literature in the field of health geography and environmental epidemiology.  Spatial analysis and health geography trace their roots back to $1854$ London when John Snow famously drew maps with markers of health incidents to locate the source of a cholera outbreak~\cite{snow1855mode}. 
Spatial epidemiology has ever since contributed to our understanding of how diseases spread across geographies~\cite{gatrell1996spatial} with roots to spatial statistics~\cite{fotheringham2000quantitative}. 
The state-of-the-art in statistical epidemiology of non-infectious disease treats occurrences as a spatio-temporal point process\cite{moller1998log, diggle2007model, diggle2005point}. This family of techniques however focuses on predicting incident frequencies of a single disease and does not provide interpretations on the environmental and demographic factors that may drive the spatio-temporal occurrence of medical incidents. They trace their roots in methods such as kriging~\cite{stein2012interpolation} and they effectively reduce the problem of modeling the spatio-temporal occurrences of epidemiological incidents to a type of spatial interpolation. Works that focus on ambulance calls data and the investigation of emergency patterns geographically and temporally have been published in recent years~\cite{cusimano2010patterns, ohshige2004circadian, jones2005circadian, o2013system, peacock2006emergency}. Here we aim to build on this thread through the establishment of a stronger link between emergency incidents, human mobility patterns and urban activities as reflected in online and mobile services. 

%% file: prelim.tex
\section{Operational scope \& datasets}
\label{sec:scope}
\subsection{The North West Ambulance Service}
The work builds on data collected by the North West Ambulance Service NHS Trust (NWAS)\footnote{\url{http://www.nwas.nhs.uk/}}. NWAS is the second largest ambulance trust in England, providing services to a population of seven million people residing in a region of approximately 5,400 square miles in size. 
Paramedic crews, call operators and other members of the organisation provide around the clock accident and emergency services to those in need of emergency medical treatment and transport, responding to hundreds of thousands emergency calls per year.
Specialised and trained staff offer life-saving care to patients and take people to a hospital or other types of treatment units if required. 
Calls that result in an ambulance dispatch come primarily via the $999$ line. One of the operators in charge will ask the caller a sets of questions to identify the degree of emergency and subsequently they will assign a dispatch code to the call. Dispatch codes are numeric and correspond to a general classification of health incidents (e.g. falls, traumatic injuries, 
assault, psychiatric etc.). If the call requires a response, an ambulance vehicle unit receives the instruction, and  makes its way to the incident's location as quickly as possible.

\subsection{Datasets and their properties}
Next we describe the characteristics of the data we employ in the following sections.
The ambulance calls dataset has been collected by the North West Ambulance Service (NWAS). We then employ additional datasets to build a number of population indicators profile and activity using open government and web sources (Foursquare). 
\paragraph{\textbf{Ambulance calls dataset}}
The data is comprised of $5.6$ million calls the service has responded to from April 2013 to March 2018. Each incident is associated with a \textit{dispatch code} which corresponds to the type of the medical condition or cause that led to the call (e.g. suicide, fall, traffic incident etc). There are 35 codes ranging from 1 to 35, though other codes for rare cases may also be used.  
We use the incident number to stratify ambulance calls by nature. In Table~\ref{discodefreq} we present the corresponding \% of calls of the ten most common natures. More than $75\%$ of calls are health incidents described by these top natures. Nature 35 corresponds to interhospital facility transfers. This involves moving one patient from one hospital to the other (e.g. due to bed availability or the availability of specialised equipment for certain exam types) and calls by medical professional to hospitals amongst others. Given that code 35 does not correspond to epidemiological incidence that emerges naturally through population activity, but instead its an operational feature of the service provided, we have excluded this code from our analysis. Each incident is time stamped with the hour the call took place. Additionally the region the call originated from is noted. In term of geographic accuracy, the administrative level of \textit{lower super output areas} (LSOAs) in the UK is used. We describe the characteristics of those in detail next.
\paragraph{\textbf{Spatial boundaries, population descriptors and socio-economic indicators}}
As noted the UK Lower Super Output Areas (LSOAs) is the unit of geographic aggregation considered in this work. Output areas were originally created such that populations were approximately similar in terms of demographic and socio-economic makeup as well as size~\cite{cockings2011maintaining}. LSOAs were designed to maintain population homogeneity with a target population size of around 1500. This number typically fluctuates but with population levels being centered around that average value. Next, to predict ambulance call volumes, we use information on the number of people residing in each LSOA, but also information about \textit{daytime population} which is simply sum between the number of workers employed at a given area and the number of residents belonging to age groups below 16 and above 70 year old.
Finally, in terms of socio-economic indices that characterise an area's population, we employ the \textit{Index of Multiple Deprivation} (IMD)~\cite{payne2012uk} which is a widely used score calculated by the UK government and obtained through the compilation of a set of deprivation and quality of life indicators. These include income ($22.5\%$), employment deprivation ($22.5\%$), crime levels ($9.3\%$), accessibility to education ($13.5\%$), health deprivation and disability ($13.5\%$), barriers to housing ($9.3\%$) as well as the quality of the living environment ($9.3\%$) where in parenthesis we note the weights of the constituent variables in determining the IMD score for an area. 
\paragraph{\textbf{Spatial distribution of calls}}
\begin{figure}[t]
\centering
\includegraphics[scale=0.24] {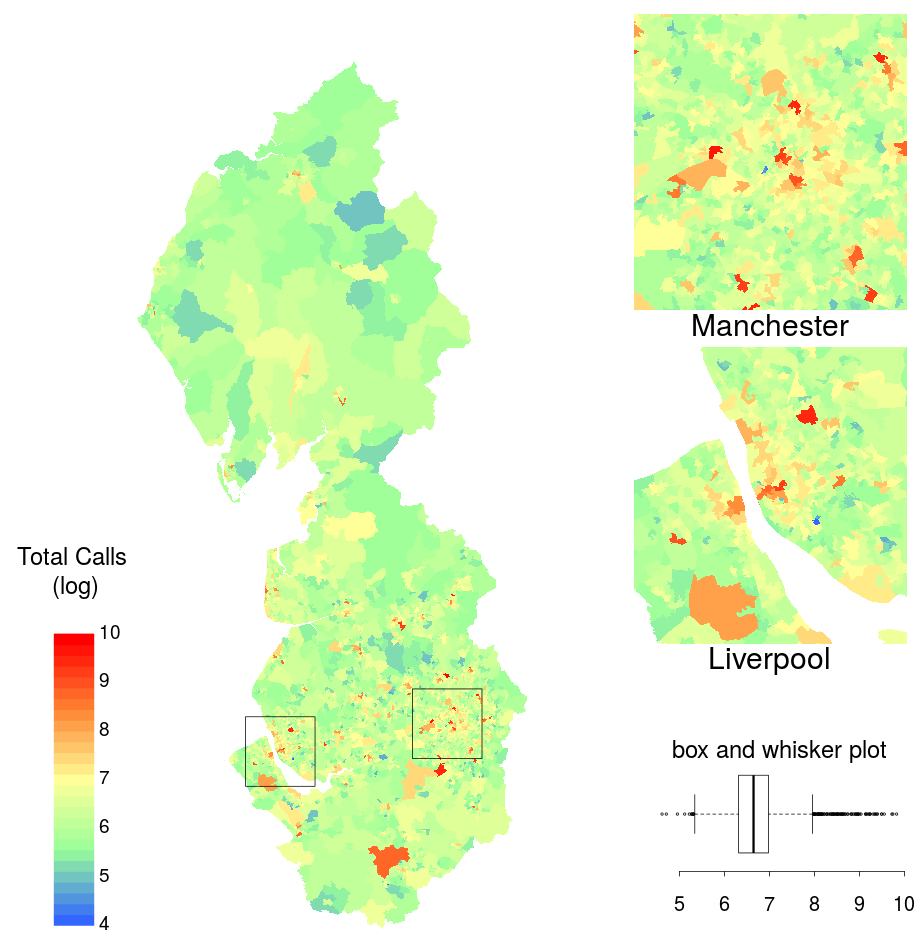}
\caption{Choropleth map for (natural log) for all calls with location of hospitals circled with zoomed  Manchester and Liverpool areas. Bottom right shows a box and whisker plot of the data. }
\label{totalcalls}
\end{figure}
One of the primary goals of the present work is to understand what population activity factors are associated to what types of ambulance calls spatio-temporally. In this context, one plausible question is how the frequency of calls is spatially distributed across regions.
In Figure~\ref{totalcalls} we plot views the frequency of ambulance calls across different LSOAs in the North West of England. The choropleth map for the total number of calls across the whole five year period is shown using the natural logarithm to represent the number of calls in each of the regions. One can notice the sizes of LSOAs varying considerably to maintain approximately similar population sizes per design intent, with urban centres corresponding to smaller geographic areas of higher population density. The squares on the main map are zoomed views of the two biggest cities, Manchester and Liverpool. As can be observed, there is a significant variation in calls across geographies with call volumes that are different by several orders of magnitude. Furthermore, calls concentrated in the two main urban centers of England's north west region in line with research literature on the subject suggesting that the vast majority of calls occur in urban environments~\cite{peacock2006emergency}. These results also suggest that knowledge about the geographic and land use characteristics of an area becomes significant when predicting ambulance calls as we do in Section~\ref{sec:prediction}.
\begin{table}[]
\centering
\begin{tabular}{lll}
\hline
Dispatch Code & Complaint                         & \%   \\
\hline
35            & Healthcare Practitioner Referral & 16.2 \\
17            & Falls                             & 13.3 \\
6             & Breathing Problems                & 10.8 \\
10            & Chest Pain                        & 9.1  \\
31            & Unconscious/Fainting              & 7.6  \\
26            & Sick Person                       & 7.1  \\
12            & Convulsions/Seizures              & 4.7  \\
25            & Psychiatric/Suicide               & 4.0  \\
23            & Overdose/Poisoning                & 3.0  \\
30            & Traumatic Injuries                & 2.5  \\
\hline
\end{tabular}
\caption{Ten most frequent dispatch codes ranked by frequency.}
\label{discodefreq}
\end{table}
\paragraph{\textbf{Population mobility and activity patterns.}}
Finally, we employ a dataset collected through the location intelligence platform Foursquare. This is comprised of check-ins pushed on Twitter, collected over 11 months in 2011. For every check-in information about the place the user has checked-in becomes 
available, including its location in terms of latitude and longitude coordinates. Moreover,
for every Foursquare venue we know the category of it (Coffee Shop, Italian Restaurant etc.). considering more than 300 categories included in the Foursquare database \footnote{\url{https://developer.foursquare.com/docs/resources/categories}}. In this work we employ Foursquare categories as a proxy to urban activities and land use.  In total, in the area that falls under the operational scope of the North West Ambulance Service, we have observed approximately $240$ thousand check-ins over the $11$ month period considered. 
We have matched every Foursquare venue in our dataset with an LSOA through a spatial join operation between the venues dataset and the polygons describing the boundaries of the LSOAs. 

%% file: analysis.tex
\begin{figure}
	\centering
	\includegraphics[width=0.47\textwidth]{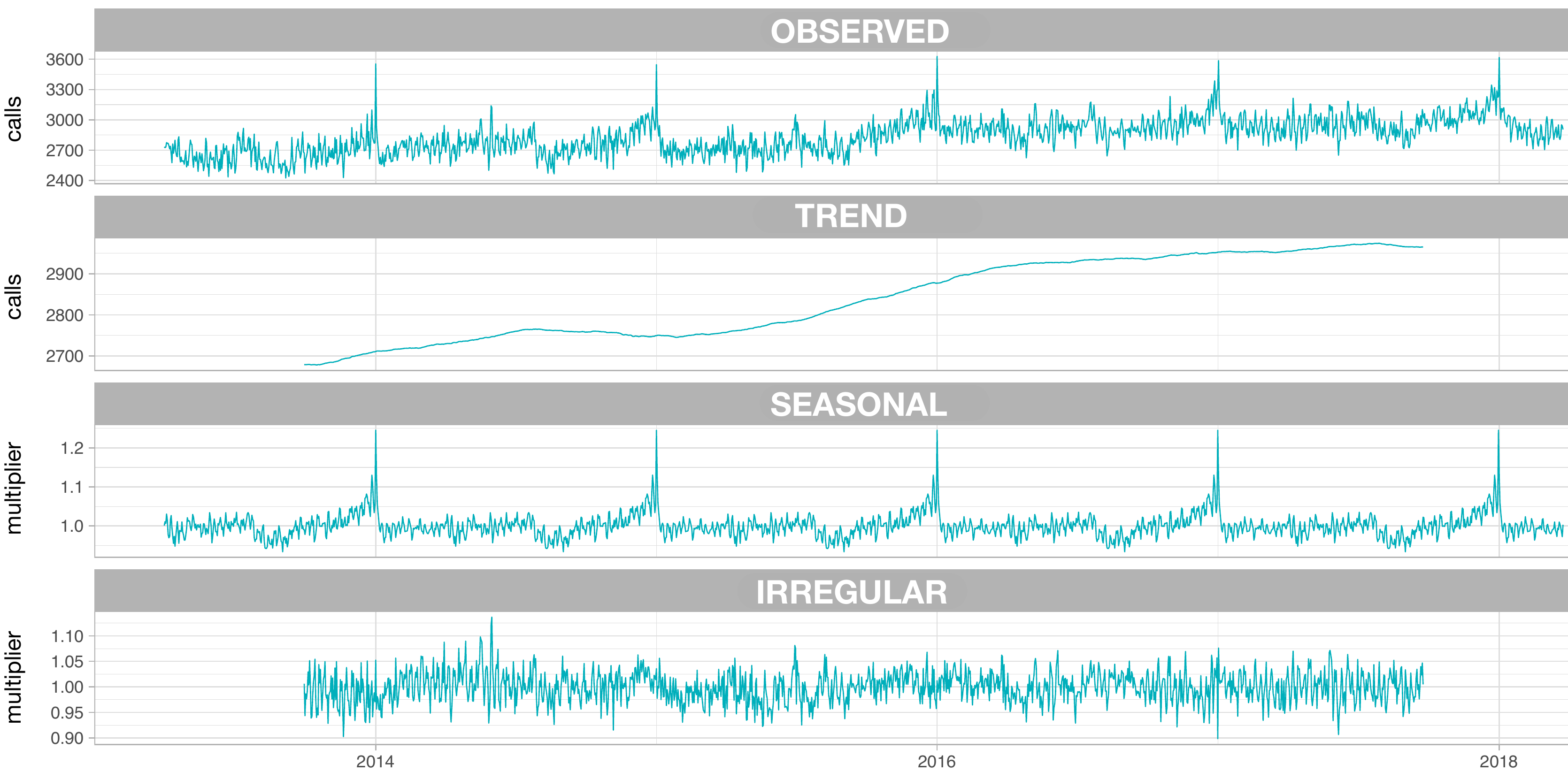}
	\caption{Decomposition of the daily time series of ambulance calls shown in four panels: (a) oringal call frequency data, (b) overall longitudinal trend, (c) seasonality patterns and (d) evolution of irregular terms after removing the seasonality and trend components from the time series. }
	\label{fig:decomp}
\end{figure}
\section{Temporal Dynamics of Ambulance Calls}
\label{sec:analysis}
In this section we conduct an analysis on the temporal dynamics of ambulance calls. First we perform a time series decomposition~\cite{prado2010time}, aiming to understand the basic properties of the data in terms of long term time trends and seasonality. We then focus on the periodic diurnal and weekly cycles of call frequencies that correspond to various health conditions (natures). Our analysis shows that different natures have different temporal profiles in terms of the times of the day or days of the week that they peak or drop. However, when we control for a given nature in our experiments, we observe stable temporal patterns not only within a single geographic region, but across regions as well albeit weaker. We present our results in detail next.
\subsection{\textbf{Ambulance Call Time Series}}
Initially the ambulance calls dataset is explored purely as time series. The top panel of Figure~\ref{fig:decomp} shows the time series aggregated on a daily basis. It can be seen that there are five peaks on regular intervals, which indicates yearly seasonality. There is a significant surge in the number of calls received by the ambulance service during the Christmas and New Years Eve period due to the increased population activity and engagement with, at times risky, entertainment activities during the festivity period in the UK. In order to capture the overall series trend which can be seen on the second panel in Figure~\ref{fig:decomp}, centered moving averages (CMA) of call frequencies are calculated with the length of the seasonality set to 365 days. We note the steep increase in the number of calls during 2015. This has been largely driven by the introduction of the so called 111 service~\footnote{\url{https://en.wikipedia.org/wiki/NHS_111}} in the region that the North West Ambulance Service operates. This has been a new UK wide service call line targeting incidents that require less urgent attention and with the aim to allow for better resource allocation and treatment of more serious incidents. There are two ways of calculating the seasonality, depending on the type of relationship between trend and seasonality in the time series. Both additive and multiplicative compositions are performed and the current dataset showed to be multiplicative. Seasonality in multiplicative time series is calculated by dividing each data point with the corresponding value from the CMA. The dataset is five years long, so each day of the year is repeated five times. Then seasonality, displayed on the third panel of Figure~\ref{fig:decomp} is calculated by taking the average of these five values. 
The final panel of Figure~\ref{fig:decomp} displays the irregular terms after trend and seasonality are removed from the data. Naturally, variations do exist with regards to the baseline time series model as population activity can present irregular patterns as well. 
\subsection{\textbf{Diurnal and Weekly Cycles in Ambulance Calls}}
\begin{figure}
\begin{subfigure}{0.23\textwidth}
\includegraphics[width=\linewidth]{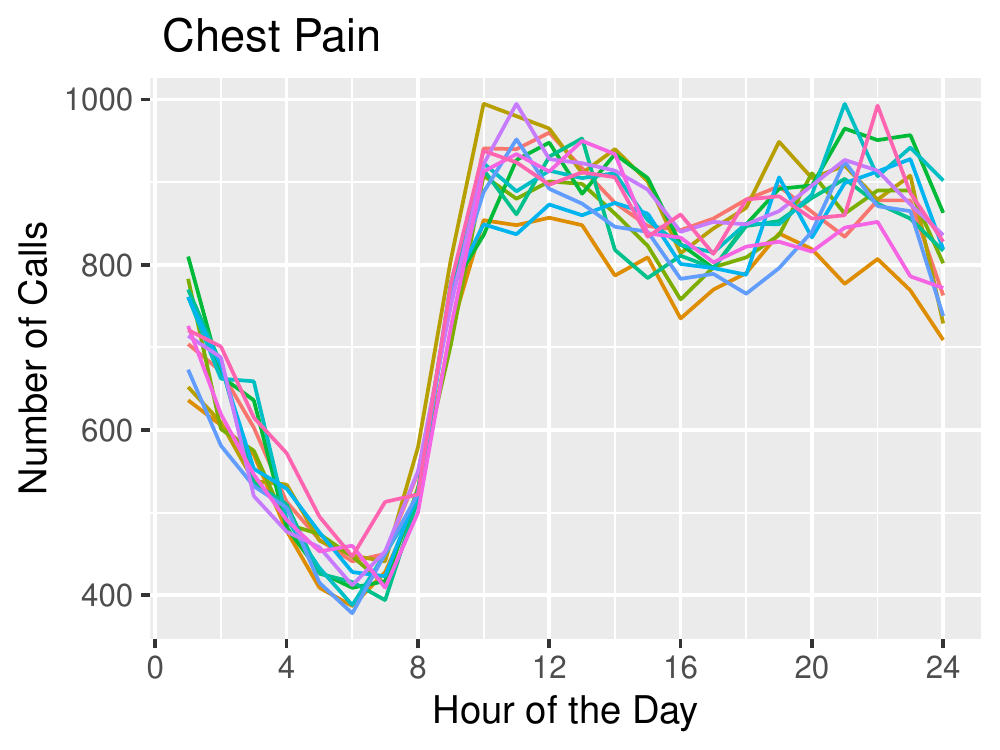}
\end{subfigure}
\begin{subfigure}{0.23\textwidth}
\includegraphics[width=\linewidth]{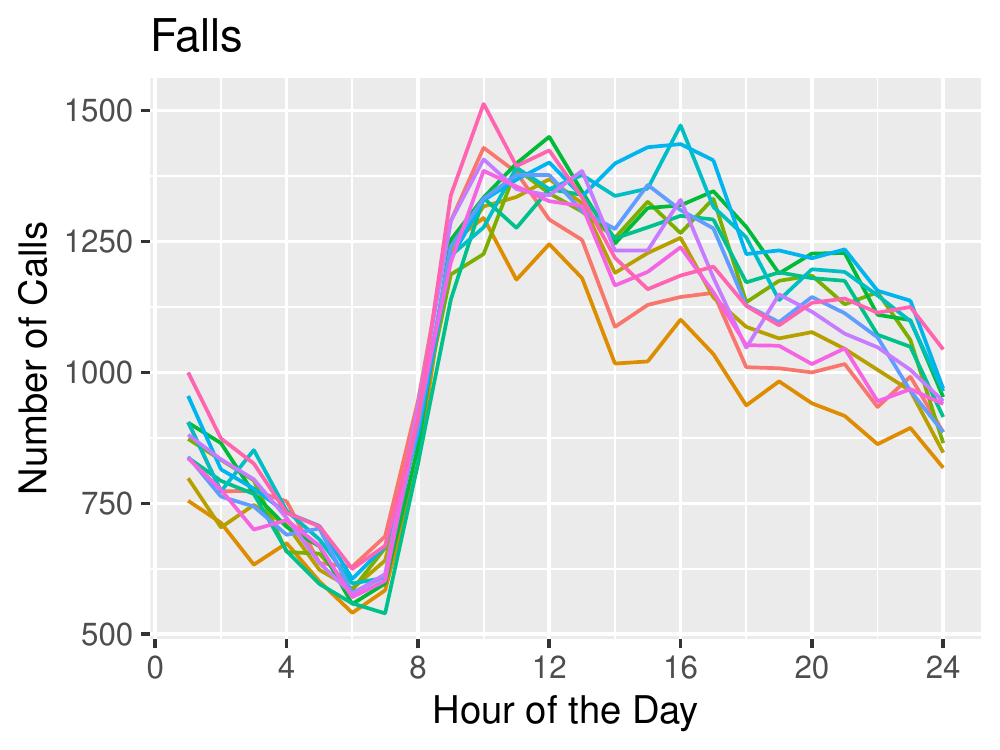}
\end{subfigure}
\begin{subfigure}{0.23\textwidth}
\includegraphics[width=\linewidth]{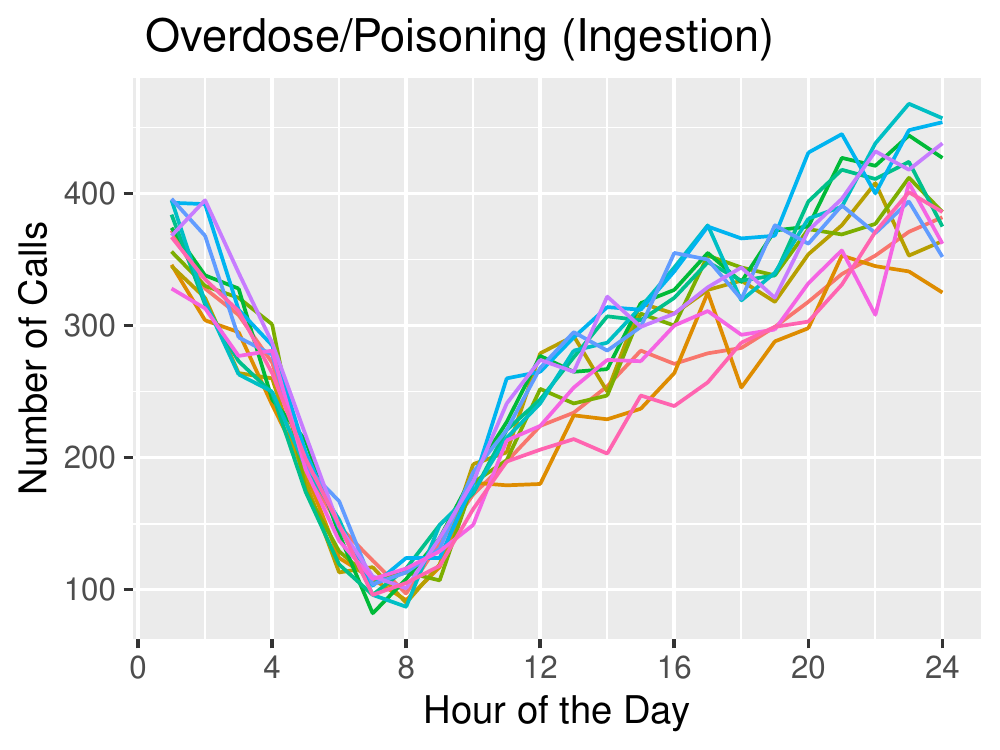}
\end{subfigure}
\begin{subfigure}{0.23\textwidth}
\includegraphics[width=\linewidth]{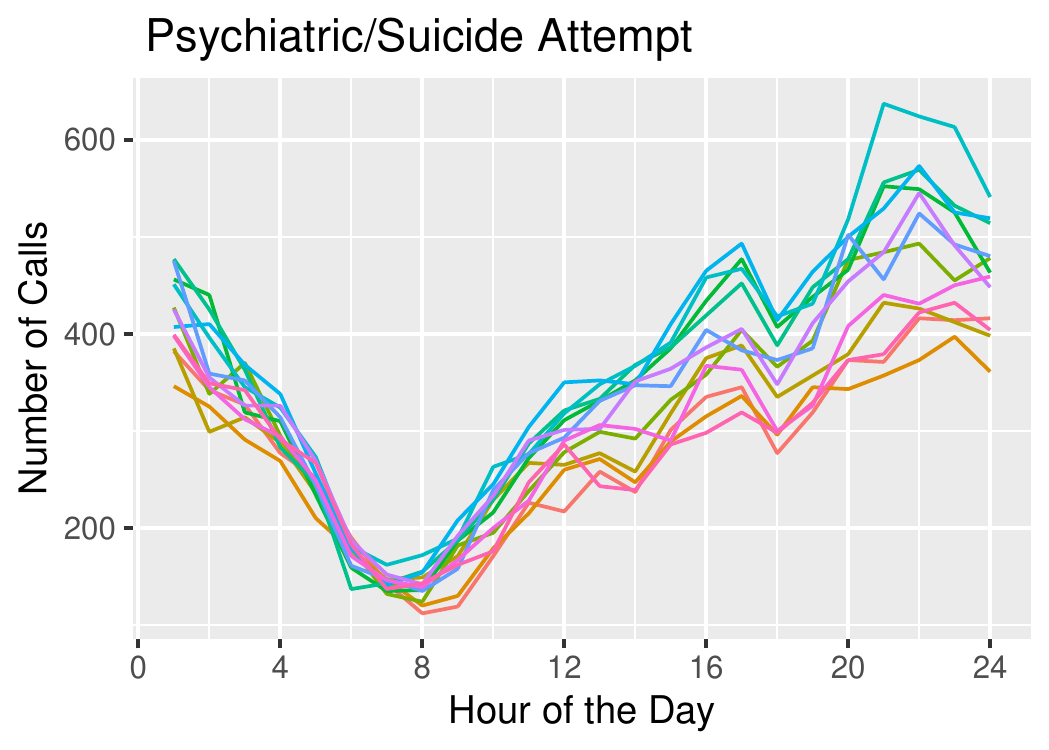}
\end{subfigure}
\begin{subfigure}{0.23\textwidth}
\includegraphics[width=\linewidth]{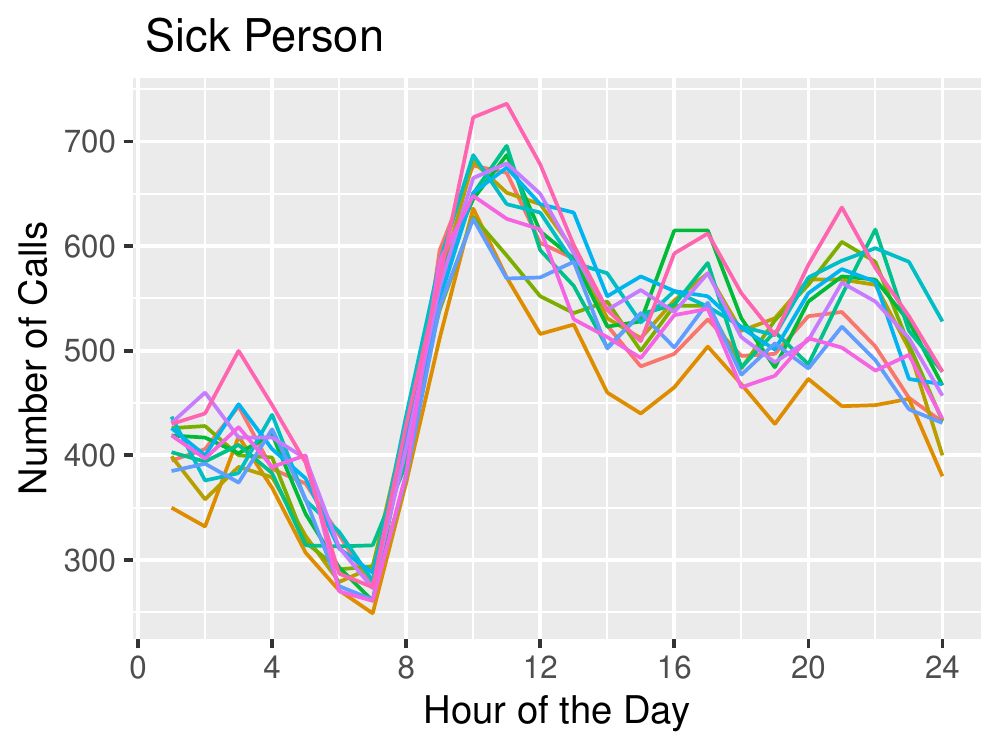}
\end{subfigure}
\begin{subfigure}{0.23\textwidth}
\includegraphics[width=\linewidth]{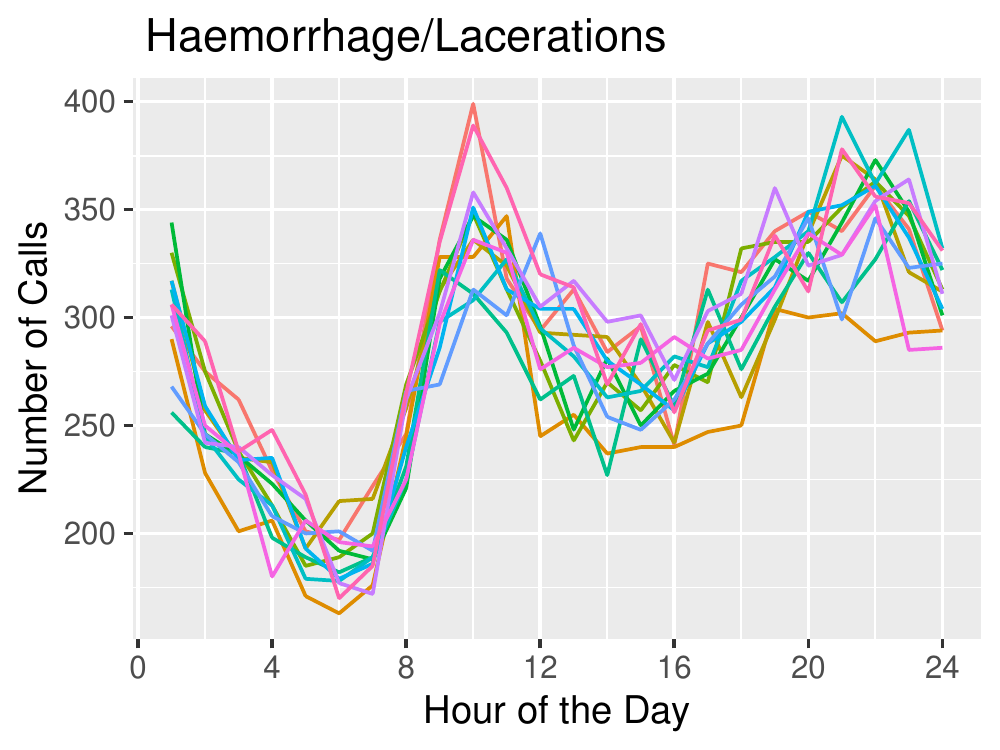}
\end{subfigure}
\caption{Diurnal temporal profiles six six natures: in each case we draw twelve curves, one for each of the twelve months of a year.} \label{fig:monnat}
\label{fig:diurnal}
\end{figure}

Next, we aggregate ambulance calls to obtain daily and weekly frequency representations for each of the natures considered. In Figure~\ref{fig:diurnal} we present the diurnal patterns for 6 natures: Chest Pain, Falls, Overdose/Poisoning, Psychiatric/Suicide Attempt, Sick Person and Haemorrhage/Lacerations. In each case, we consider a monthly time window of ambulance call events for each of the five years in the dataset and generate a daily frequency curve based on the data from that month. We repeat the aggregation process for twelve different months in the dataset and plot the corresponding curves. We make two key observations. First, each nature appears to be characterised by a stable temporal profile. Variations exists across data collected across different months, but the shape of the curve remains relatively stable. Second, different natures can feature significantly different profiles. For instance Falls, Sick Person and Falls are more common during the morning hours, peeking around 9-10am, whereas Overdose/Poisoning and Psychiatric/Suicide Attempt cases become progressively more common as we approach nighttime. Incidents that fall in the Haemorrhage/Lacerations are bimodal and surge twice a day with more pronounced peak in the morning and a smaller one during the evening hours.
 
\begin{figure}
\begin{subfigure}{0.4\textwidth}
\includegraphics[width=\linewidth]{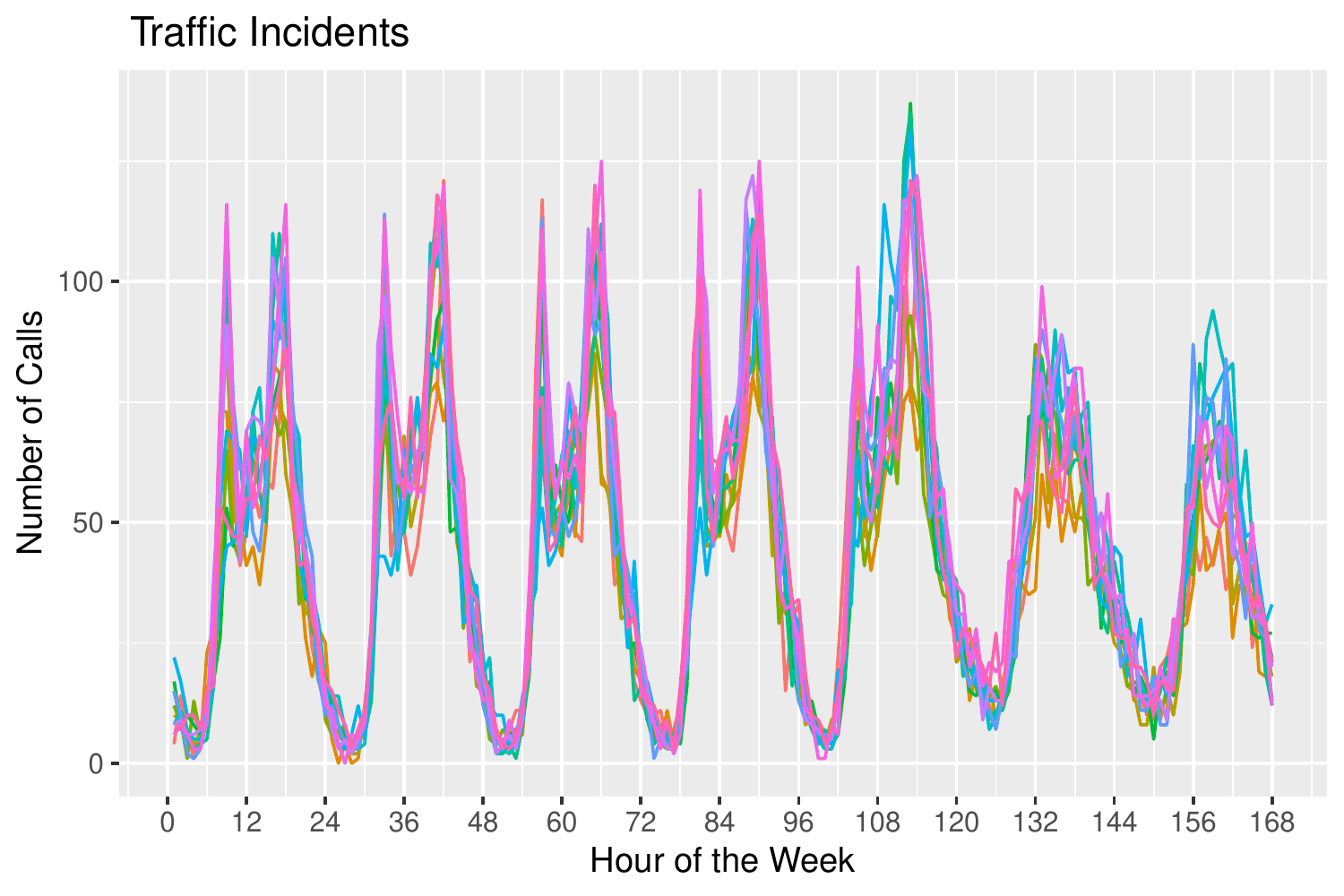}
\end{subfigure}
\begin{subfigure}{0.4\textwidth}
\includegraphics[width=\linewidth]{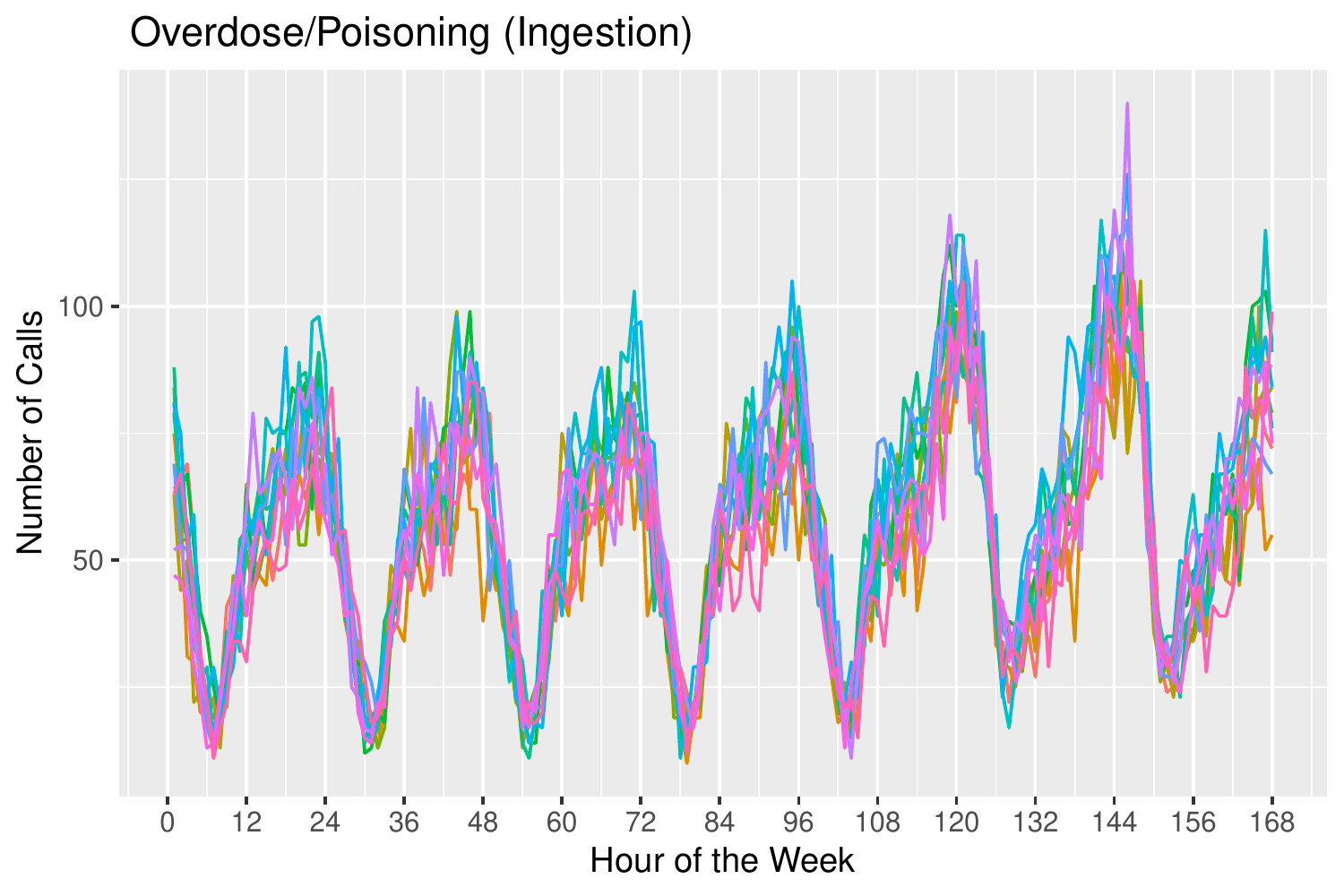}
\end{subfigure}
\caption{Weekly representations of ambulance calls frequencies for two natures: Traffic Incidents and Overdose/Poisoning. Hour 0 represents the Monday 0-1am time window and 168 hours in the week are considered.} 
\label{fig:weeklycycles}
\end{figure}

What is more, when we investigate the temporal frequency patterns of incidents considering weekly representations, as shown in Figure~\ref{fig:weeklycycles}, we note that patterns vary across natures in terms of how they become more or less prominent at different days of the week. Traffic incidents are more common during weekdays and peak during travel peak times, whereas Overdose/Poisoning incidents become significantly more common during weekends as people are more likely to intoxicate themselves during nightlife activities. Overall the characteristic diurnal and weekly cycles in ambulance calls resemble typical patterns observed in ecological systems as they naturally reflect periodic changes in collective behaviour. In Section~\ref{sec:urbanrisk} we will establish a strong link between population activities and emergency incidents. Next, we study how the temporal profiles of different natures vary across different geographic regions. 
\subsection{Geographic Variations in Temporal Dynamics}
\begin{figure}[ht!]
\begin{subfigure}{0.5\textwidth}
	\centering
	\includegraphics[width=\linewidth]{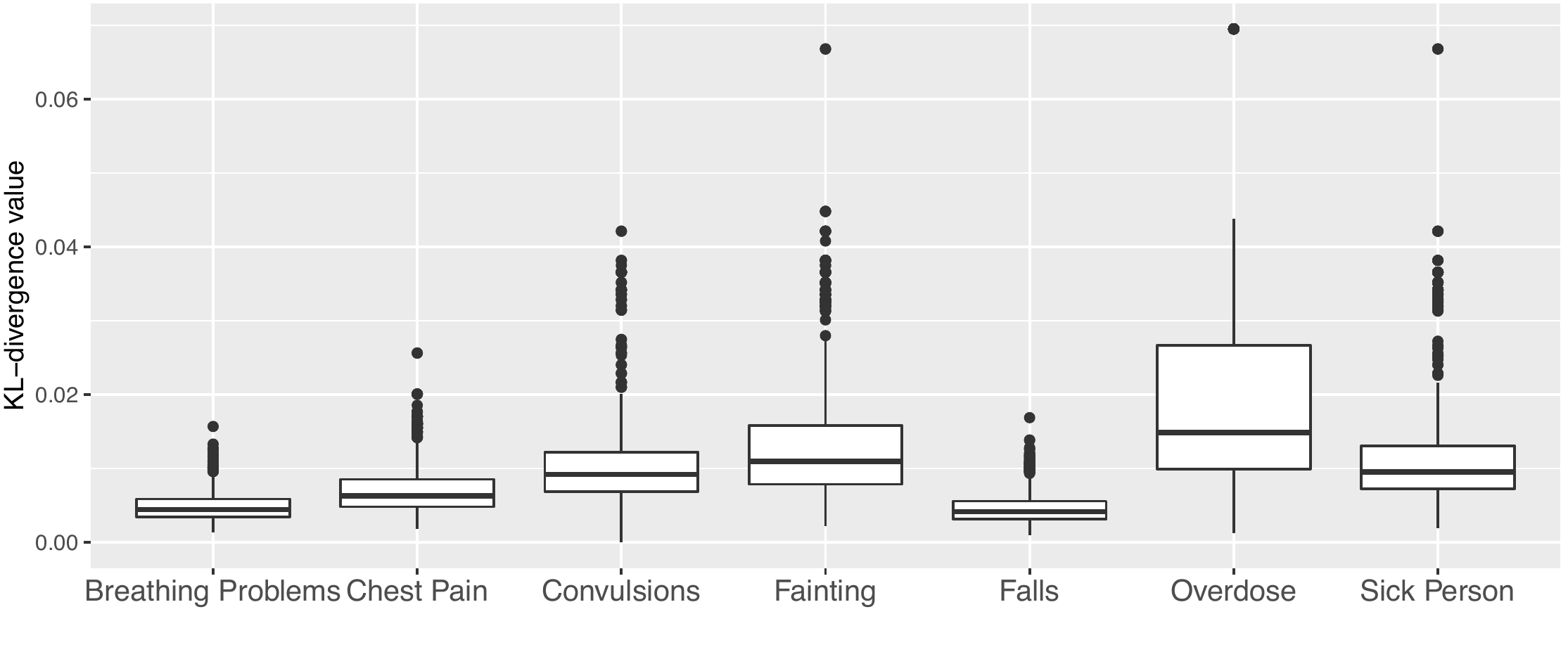}
	\caption{Same region comparison}
	\label{fig:kldivwithin}
\end{subfigure}
\hspace*{\fill}
\begin{subfigure}{0.5\textwidth}
	\centering
	\includegraphics[width=\linewidth]{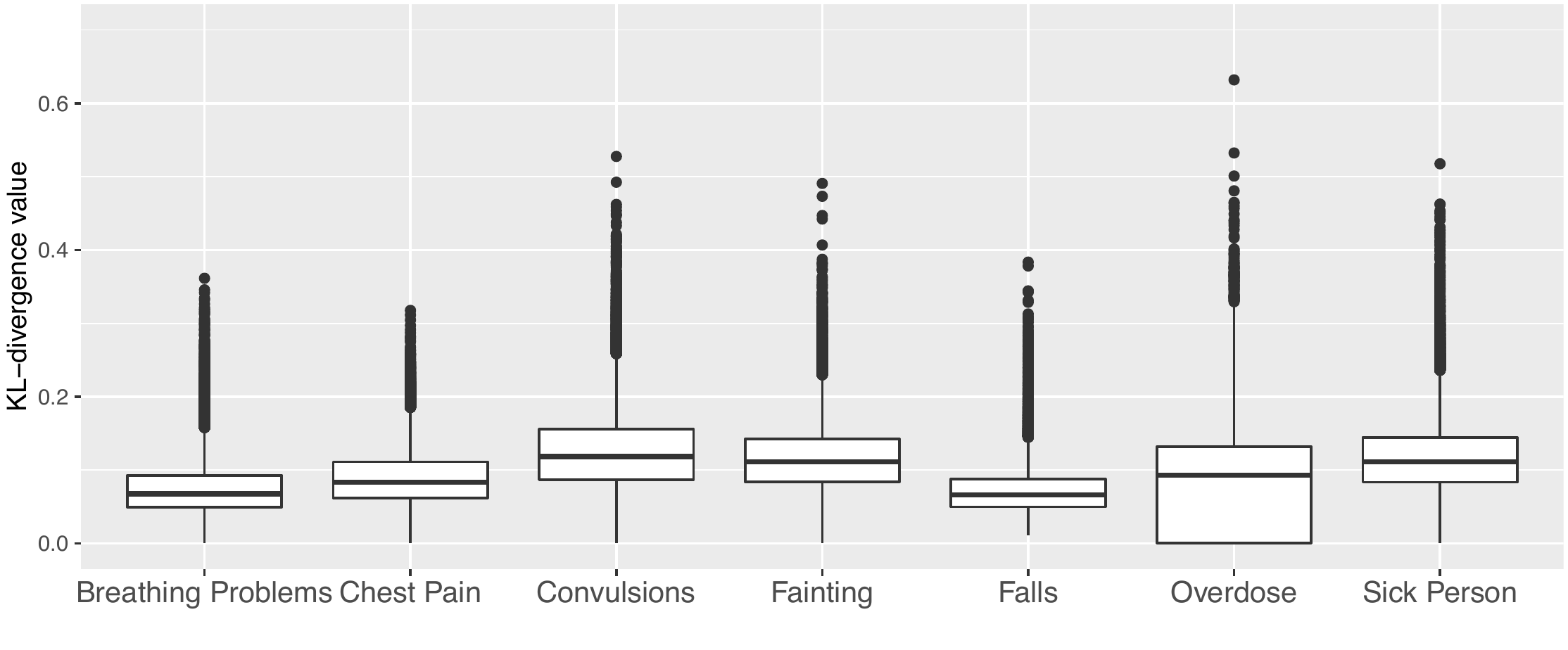}
	\caption{Cross region comparison}
	\label{fig:kldivwithout}
\end{subfigure}
\caption{KL-divergence scores obtained by comparing successive montly time windows of weekly aggregations for different natures.}
\end{figure}

In Figure~\ref{fig:kldivwithin} we quantify the observations of discussed in the previous paragraph and show the KL-divergence scores obtained from the weekly frequency patterns of calls for different natures, within the same geographic region. As we can see the weekly call frequencies feature low divergence scores with some incidents types like Overdose/Poisoning to present higher variance scores. In Figure~\ref{fig:kldivwithout} we perform the same analysis but we now measure the KL-divergence of weekly frequency curves across different geographic regions. That is, we pick a random pair of regions in the dataset and considering a nature we measure the divergence for calls that happened in the same time period. While a similar picture is drawn in terms of the differences across natures, cross region comparisons demonstrate higher divergence scores. This is somewhat expected as we consider different geographic pools of acting populations. Lifestyle and population activities are somewhat homogeneous in a county, but variations can exist across geographies due to the different socio-economic make-ups of local populations amongst other factors. Reflecting on these observations, in Section~\ref{sec:prediction}, when we predict ambulance calls over space and time, we build region specific features both that are characteristic of the resident population, but also features that capture the temporal activity dynamics of a specific locale.

%% file: urbanrisk.tex
\section{Mapping Urban Activities to Risk for Emergencies}
\label{sec:urbanrisk}

In the previous section we demonstrated how the frequency of ambulance calls corresponding to a given a nature features characteristic temporal signatures manifesting in the course of daily and weekly cycles. Moreover, certain types of emergency incidents appear to follow similar temporal trends with the activities that populations engage with at local areas. 
Next, we take advantage of digital geo-tagged datasets describing population mobility and human activities over time and space to build a computational framework that captures the relationship between human activities and emergencies at scale. We demonstrate that appropriately leveraging on the spatio-temporal co-occurrence patterns of emergencies, and visits to certain place types in the city, accurate mappings can be obtained, that inform on which types of human activity pose an elevated risk for incidents that require urgent medical attention.  Furthermore, building on our findings we devise a novel metric, referred to here as the \textit{urban activity risk} of a geographic area. The metric aims to capture the potential for an area to yield a medical incident.

\subsection{Spatial Attractiveness between Place Types and Medical Natures} 

To begin with we model the co-occurrence patterns of Foursquare place types at a geographic area and various types of medical incidents. We assume that if a place type and medical incident co-occur frequently across space, then they tend to \textit{attract} each other more. In other words, we expect a certain type of emergency to be more likely to take place at an area where a particular set of place types becomes popular. The assumption is founded on previous literature that has established that human activities across space are not randomly distributed, but instead, they tend to form characteristic congregation patterns which are indicative of the urban functions manifesting in the corresponding local neighborhoods~\cite{marcon2009measures, brulhart2005account}. Past work has also shown how these land use patterns are indicative of the quality of urban space in terms of the retail success of local businesses~\cite{jensen2006network}. Formally, we define that \textit{spatial attractiveness}, $SA_{ij}$, of a given nature $i$ to a given place type $j$ as
\begin{equation}
SA_{ij}:=  \frac{n_i}{n_j} / \frac{N_i}{N_{all}} 
\label{eq:space}
\end{equation}
where $N_i$ is the number of times a certain incident type (e.g. Poisoning) occurs across all considered geographic areas $N_all$ and $n_i$ is the frequency of nature $i$ at areas where place type $j$ (e.g. Pub) is observed. In other words the mathematical relationship above expresses the tendency of an incident type $i$ to occur at areas where $j$ is present against the expectation of observing $i$ at a randomly selected area, irrespectively of whether place type $j$ is present or not. 
\subsection{Temporal Synchronicity between Place Visits and Emergencies} 
\begin{figure}
\begin{subfigure}{0.5\textwidth}
	\centering
	\includegraphics[width=\linewidth]{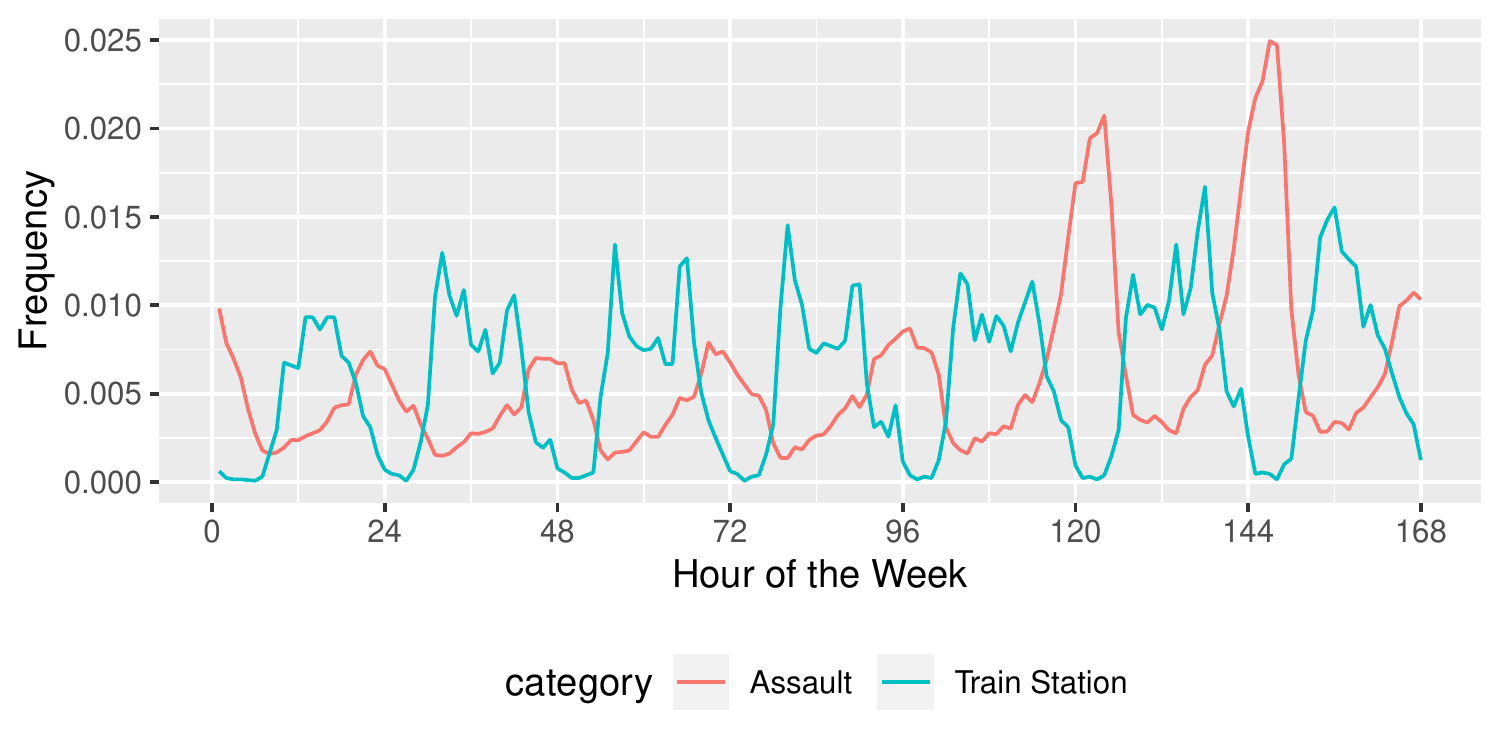}
	\label{fig:assaulttrain}
\end{subfigure}
\begin{subfigure}{0.5\textwidth}
	\centering
	\includegraphics[width=\linewidth]{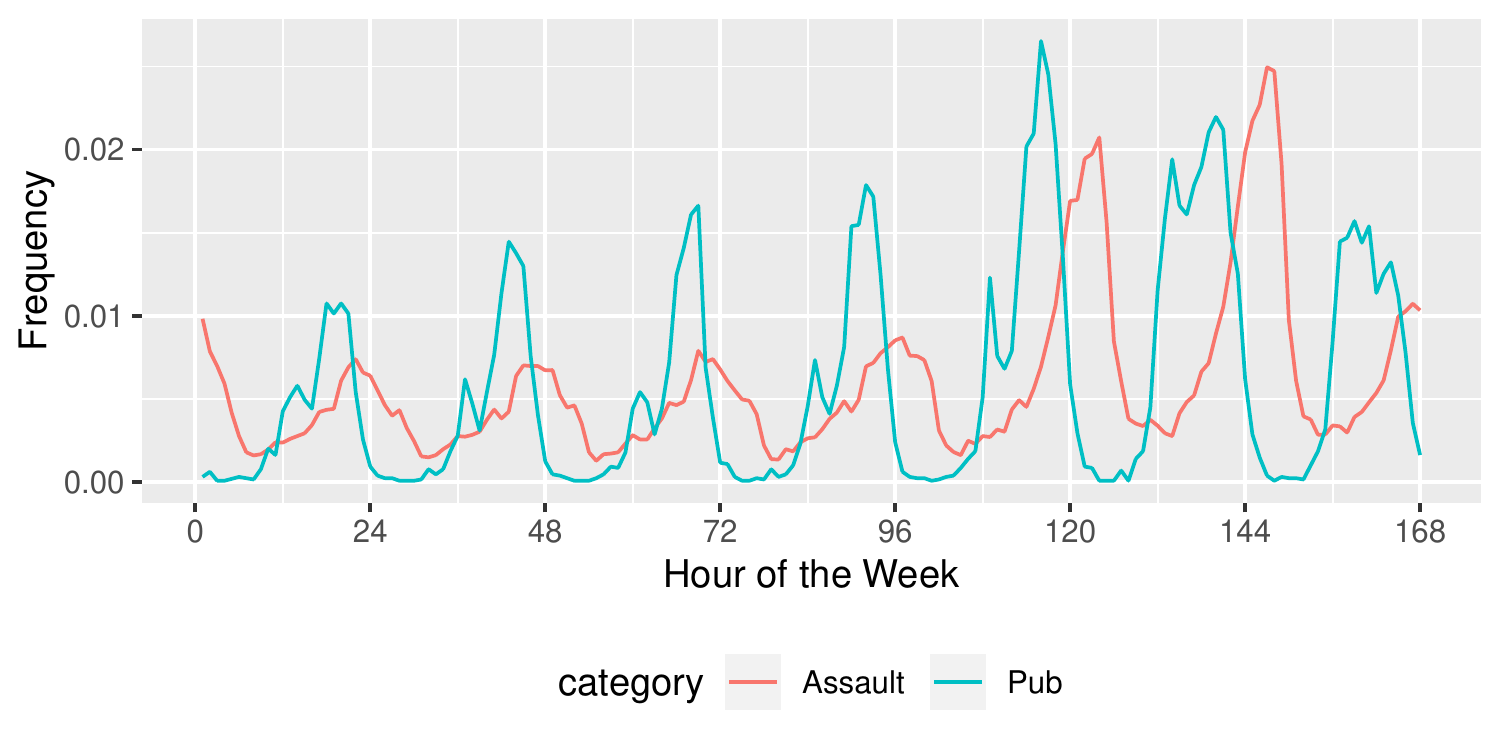}
	\label{fig:assaultpub}
\end{subfigure}
\begin{subfigure}{0.5\textwidth}
	\centering
	\includegraphics[width=\linewidth]{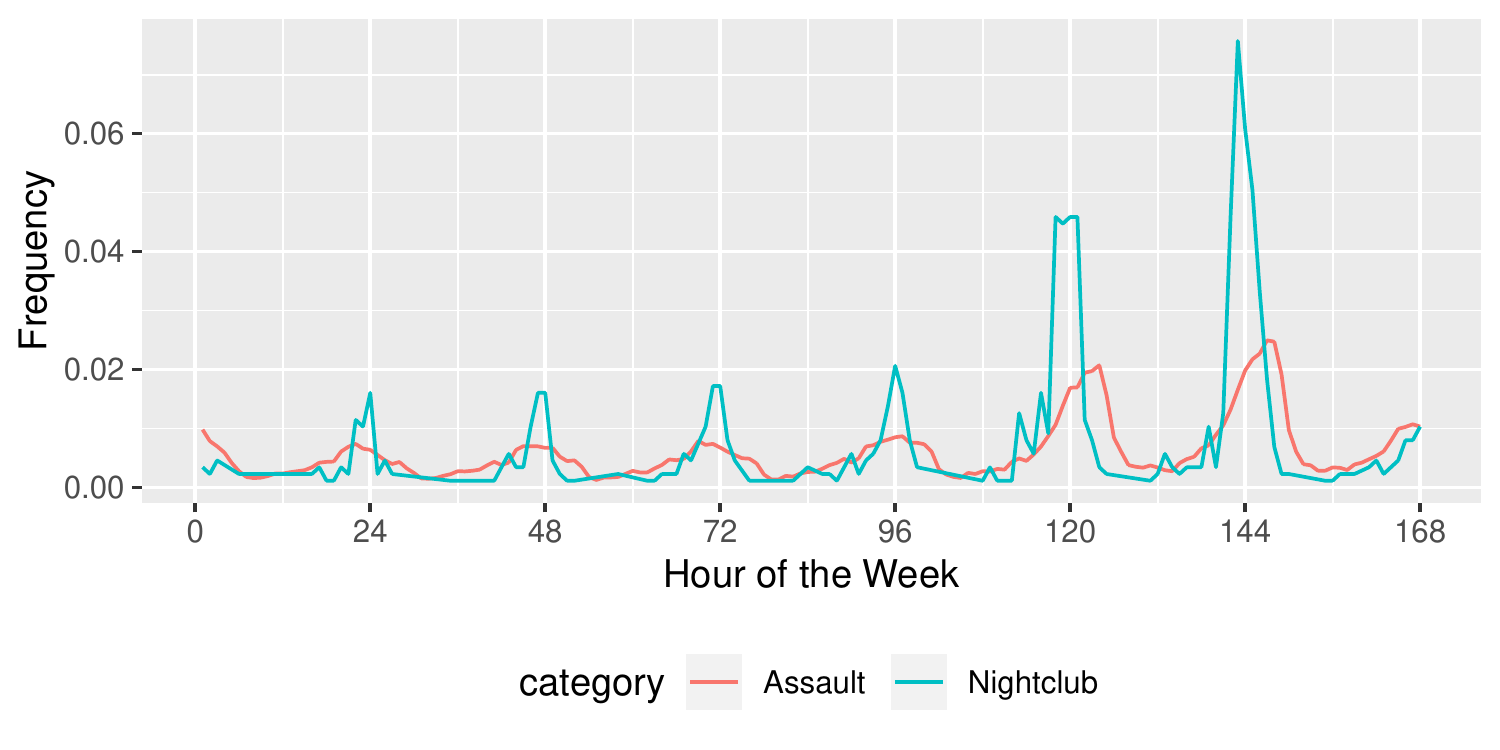}
	\label{fig:assaultclub}
\end{subfigure}
\caption{Normalised urban activity frequencies versus ambulance call frequencies. Check-ins at nightlife places such as Nightclubs or Pubs present a stronger alignment over time with Assault related ambulance calls when compared with check-in activity at Train Stations. Hour 0 represents Monday 0-1am.} 
\label{fig:activityvsnatures}
\end{figure}

Of course, an emergency incident can occur at an area where a given place is positioned, but the time it occurred might not necessarily be overlapping with the working hours of that place. Urban activities may mix spatially, but can be unsynchronised in temporal terms.
For example, a nightclub can be near a corporate office at a central city neighborhood but their hours of operation may not overlap. The metric of spatial attractiveness alone would not capture this phenomenon as it manifests over time. To clarify further this point, in Figure~\ref{fig:activityvsnatures} we compare assault categorised ambulance calls with activity at train stations, pubs and nightclubs respectively. The daily peaks of activity at pubs and nightclubs are more synched with incidents related to violent behaviour (assault). Additionally, as nightlife activity rises during the weekends, and so does the frequency of assault incidents. As it can be observed on the top panel of the same figure, the same does not hold regarding mobile user activity at train stations, which is more associated to commuting patterns. 

We quantify the above observation by extending the notion of attractiveness between place types and natures considering the time dimension. We do so by operating on the corresponding temporal frequency vectors. Formally, we consider the temporal vector $\vec{t_i}$ of a nature $i$, as the 168-dimensional vector, where each dimension corresponds to the normalised frequency of that nature at one of the 168 hours of a week. Similarly, we define as $\vec{t_j}$ the temporal signature vector of a place type $j$, considering the corresponding check-in frequencies, and then compute the cosine similarity scores across $\vec{t_i}$ and $\vec{t_j}$ to define the \textit{temporal attractiveness} score $TA_{ij}$ as:
\begin{equation}
TA_{ij} := cos(\vec{t_i},\vec{t_j})
\label{eq:time}
\end{equation}
We can then define the \textit{spatio-temporal attractiveness} of a nature $i$ to an activity type $j$, $ST\_Risk_{ij}$, which aims to capture the evelevated risk for an emergency type associated with a local activity, by combining the two similarity metrics provided in Definitions~\ref{eq:space}~and~\ref{eq:time} above by taking their product: 
\begin{equation}
ST\_Risk_{ij} := SA_{ij} \times TA_{ij}
\end{equation}
\begin{figure*}[t]
\centering
\includegraphics[scale=0.7] {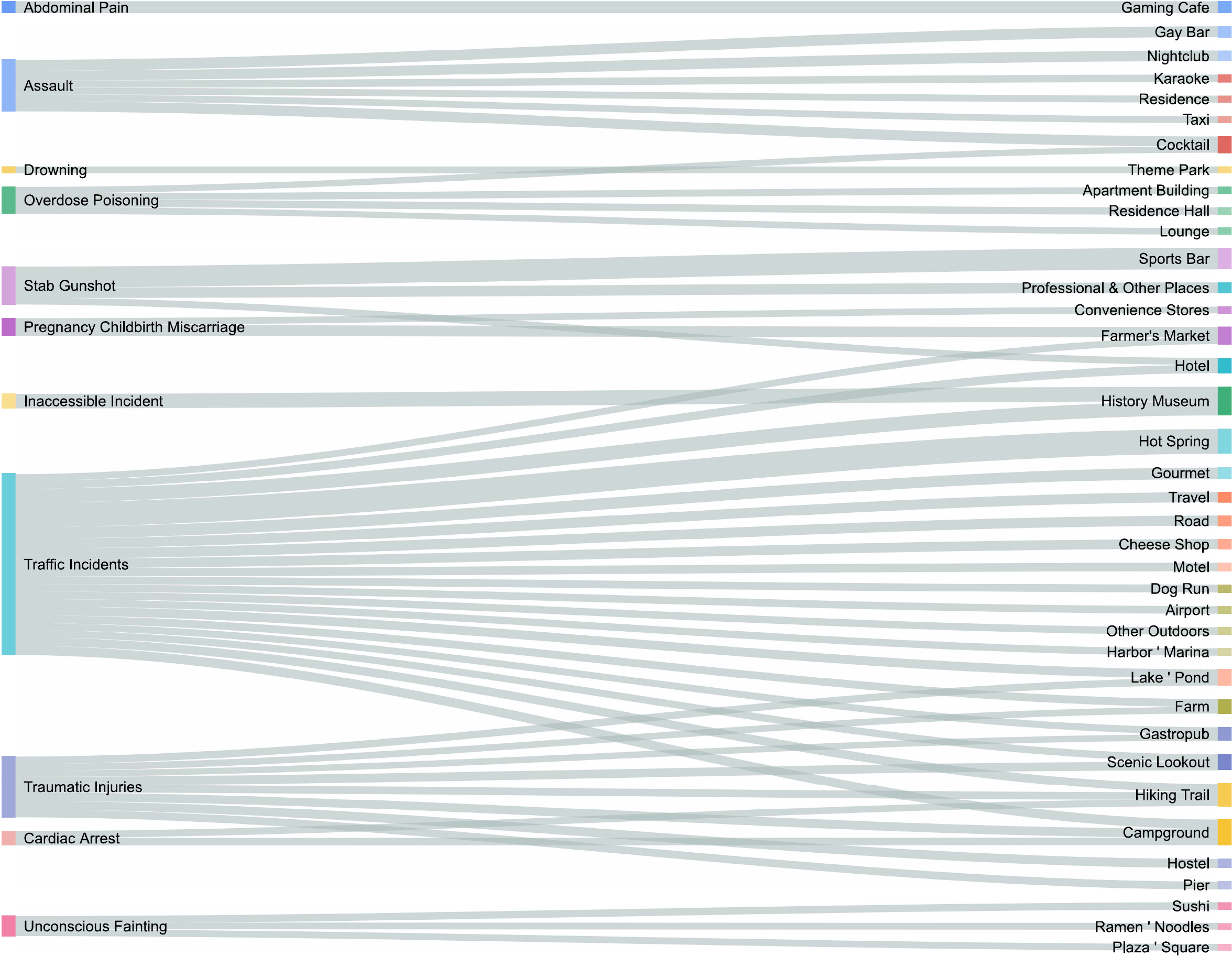}
\caption{Sankey diagram describing the relationships between ambulance call types and urban activities. Only relationships corresponding to risk scores $ST\_Risk_{ij}$ whose strength is bigger than 1.2 are included. The number shown next to each nature/activity summed \textit{risk mass} attributed to it.}
\label{sankeyattraction}
\end{figure*}
In Figure~\ref{sankeyattraction}, we present the strongest relationships between incidents and urban activities using a Sankey flow diagram  filtering out those that have an $ST\_Risk_{ij}$ smaller than $1.2$. Some of the relationships are striking, e.g. Stab and Gunshot are tightly connected to the activity at Sports Bars, when Assault is more likely at nightlife places including Nightclubs, Gay Bars, Cocktail places and Jazz Clubs. Some other incident types such as Traumatic Injuries are connected to a broader range of activities, as they are tight to population mobility and therefore are omnipresent. 

While these results present by no means a causal proof of an incident resulting due to the presence of particular type of place at an area, they can provide a general overview of what population activities are more likely to be spatio-temporally aligned with certain types of incidents and offer guidance towards a better understanding of population health informing policy in healthcare. A plausible question that rises in this context is to what extent the inclusion of the spatial and temporal components plays a role in the attraction scores obtained. In Tables~\ref{assault}~and~\ref{animal} we list the top-10 most attracted place types to Assault and Animal Bite natures. In each case we show the corresponding list when maintaining only either the spatial or temporal component of the similarity equation only. In the case of Assault incidents, the time component seems to suffice in terms of picking up activities that are intuitively associated with violence as these exceptionally take place late at nighttime. When looking at Animal Bites however, which tend to occur at outdoor spaces, the temporal component alone is not enough to explain incident-activity associations. Incorporating both the spatial and temporal dimension in the calculation of the metric allows for more relevant place type cases to emerge including: Field, Neighborhood, Campground and Marina, all outdoor places where humans can encounter animals, raising the risk for an animal bite incident.
~
\begin{table}[]
\centering
\begin{tabular}{lll}
\hline
Spatio-Temporal & Spatial Only                         & Temporal Only  \\
\hline
Jazz Club            & Residence & Gay Bar \\
Cocktail            & Cuban                             & Nightclub \\
Gay Bar             & Dim Sum                & Home \\
Nightclub            & Vietnamese                         & Cocktail  \\
Rockclub            & Tourist Information              & Bar \\
Casino            & Burritos                       & Rock Club  \\
Brazilian            & Soup              & Casino  \\
Residence Hall            & Australian               & Residence Hall  \\
Apartments            & Brazilian               & Apartments \\
Karaoke            & Bagels               & Hotel Bar  \\
\hline
\end{tabular}
\caption{Top-10 urban activities associated with Assault incidents.}
\label{assault}
\end{table}
~
\begin{table}[]
\centering
\begin{tabular}{lll}
\hline
Spatio-Temporal & Spatial Only                         & Temporal Only  \\
\hline
Diner            & Hostel & Home \\
Train Station            & Rec Center                             & Hotel \\
Field             & Sake Bar                & Neighborhood \\
Neighborhood            & Ferry                         & Fast Food  \\
Pizza            & Stable              & Apartments \\
Tea Room            & Chicken Wings                       & Pub  \\
Harbor / Marina            & Cineplex              & Grocery Store  \\
Lounge            & Voting Booth               & Road  \\
Liquor Store           & Outdoors               & Other - Food \\
Campground            & Yoga Studio               &  Bar  \\
\hline
\end{tabular}
\caption{Top-10 urban activities associated with Animal Bite incidents.}
\label{animal}
\end{table}
\subsection{Measuring urban activity risk} 
Having calculated the risk for an emergency incident $i$, given place type $j$, it is straightforward to extend the same idea for an area $a$ and define the \textit{urban activity risk}, $UAR_a$, given the set of places types that are there. We do so by iterating over all place types observed at a geographic region and summing the individual risk scores associated considering all natures. Formally, the urban activity risk score of an area is defined as: 
\begin{equation}
UAR_a := \sum_{j \in Z_a} \sum_{i \in N} ST\_Risk_{ij}
\label{eq:urs}
\end{equation}
where $Z_a$ is the set of place types in an area $a$ and $N$ is the total set of natures. The assumption made is that the land use of a given region as captured by visits at places, that are associated with specific activities, can become an indicator of elevated risk for an emergency to be observed at an area. That is, we assume that some of the geographic variations in the frequency of ambulance calls shown in Figure~\ref{totalcalls} could be explained in terms of variations in the land use of the corresponding areas. Next, in Section~\ref{sec:prediction} we demonstrate that this assumption is plausible as urban activity risk becomes a significant predictor of ambulance calls. 

%% file: prediction.tex
\section{Spatio-Temporal Prediction of Ambulance Calls}
\label{sec:prediction}
\begin{figure*}
\begin{subfigure}{0.49\textwidth}
\includegraphics[width=\linewidth]{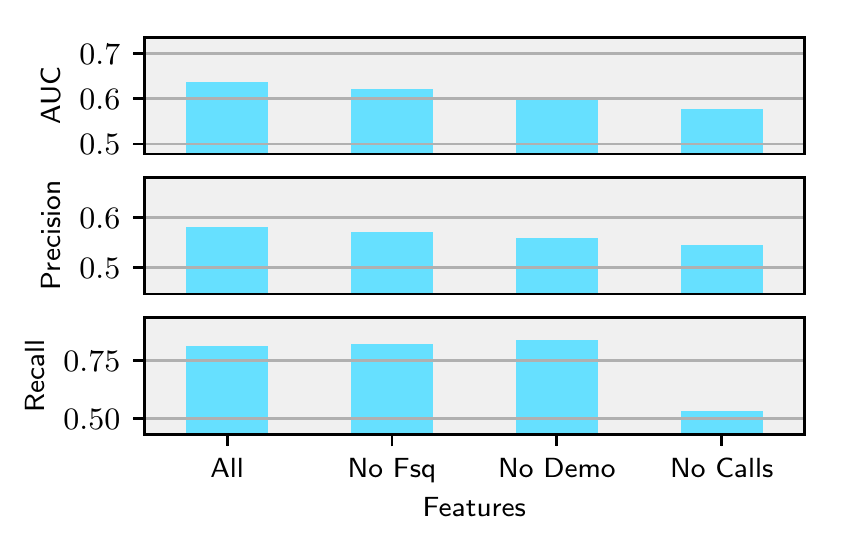}
\end{subfigure}
\begin{subfigure}{0.49\textwidth}
\includegraphics[width=\linewidth]{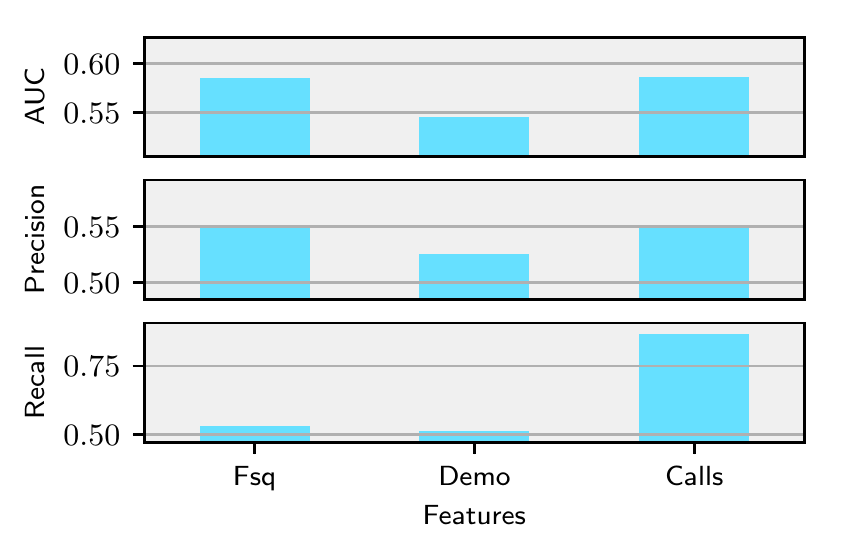}
\end{subfigure}
\caption{Model performance across three metrics (precision, recall and area under the curve). We are considering two testing scenarios. Dropping features from a given feature class (left panel) where we can observe deviations by comparing to the \textbf{All} case. Using features only from a given class (right panel) where we get a view of model performance using solely the input class of features.} 
\label{fig:modelperform}
\end{figure*}
In this Section we discuss a prediction task where the aim is predict ambulance calls at geographic areas over time. First we formalise the task, then we introduce the corresponding prediction features and subsequently present the experimental results focusing on the interpretability and operational significance of those. 
\subsection{Training a supervised learning model on ambulance call events.}
\textbf{Task formulation:}
To predict ambulance calls we train a Random Forest classifier as informally we observed a more robust behaviour in terms of prediction score stability and levels compared to linear models such as logistic regression. We use the publicly available implementation of the model available in scikit learn~\cite{scikit, breiman2001random} and have optimised the model to X number of estimators of Y maximum depth each. For the prediction problem, we define successive hourly time-windows along the time axis as prediction events assuming knowledge up to, but excluding, prediction time $t'$ . This technique is referred in time series prediction also as walk-forward validation and establishes that training and testing data are chronologically separated. For every call that took place before $t'$, we build a training example $\mathbf{x_i}~\in~X$ which encodes the values of the features of the area where the call took place and whose label $y_i$ is positive. In a similar fashion, we retrieve negative labels $y_i$ by sampling at times and areas that a call did not take place. Essentially, we are aiming to teach the model what are the crucial spatio-temporal characteristics that allow to differentiate the areas and hours where calls are likely to take place versus those that are less likely. In machine learning terms the problem is an online binary classification task, where the classifier is trained and outputs either a $+1$ (call) or a $-1$ (no call) class instance. We highlight that we define the prediction problem in geographic areas that are relatively small as LSOAs host typically $1,5$ thousand residents. At that level of spatial granularity, hourly time rates of ambulance calls are low. Class imbalance is therefore highly skewed with negative events being more common and a mean observed rate of around $0.02$, which is two calls every one hundred hours). 
\\
\textbf{Prediction features:}
In terms of prediction features, we use a number of area population descriptors as well as temporal ones. For an area $a$, we calculate the \textit{urban activity risk} score, $\mathbf{UAR(a)}$, as noted in Equation~\ref{eq:urs}, its \textit{residential population}, $\mathbf{ResPop(a)}$ as noted in the 2011 UK census survey, as well as its \textit{daytime population} $\mathbf{DayPop(a)}$ defined in Section~\ref{sec:analysis}. In terms of socio-economic descriptors, we employ the \textit{index of multiple deprivation} noted as $\mathbf{IMD(a)}$ (described also in Section~\ref{sec:scope}). We refer to this group as \textit{demographic features} and indicate them using the abbreviation $\mathbf{Demo}$.

Besides, we employ features that capture the temporal patterns of ambulance calls historically at an area. To begin with, we simply define the \textit{call\_history} of an area, $\mathbf{Hist(a,t')}$ as the total number of calls that took place there before prediction time $t'$. Then, guided by the characteristic temporal patterns of ambulance calls described in Section~\ref{sec:analysis} we define the historic frequency of calls at an hour of the day at area $a$ as $\mathbf{HoD\_hist(a,h)}$, whereas $\mathbf{HoW\_calls(a,h)}$ considers the past call frequencies at $a$ during hours of the week, with $h$ running in the discrete intervals $\{0 \dots 23\}$ in the former case and $\{0 \dots 167\}$ in the latter case. We note with $\mathbf{HoD\_hist\_f}$ and $\mathbf{HoW\_hist\_f}$ the corresponding relative fractions of calls at that hour when dividing with the total number of calls that took place thus far in the area. The $\mathbf{Day\_W\_hist(a,d)}$ measures the number of calls at day $d$ of a week, where Monday is encoded as $0$ and Sunday as $6$. We refer to this group as the $\mathbf{Calls}$ set of features. 

Finally, in addition to the urban risk score defined above, we consider the temporal activity of Foursquare users at an area: the $\mathbf{Fsq\_HoD(a,h)}$ and $\mathbf{Fsq\_HoW(a,h)}$ count the number of check-ins at an area given an hour of the day and hour of the week respectively. With $\mathbf{HoD\_Fsq\_f}$ and $\mathbf{HoW\_Fsq\_f}$ we note the corresponding fractions when dividing the two aforementioned features with the total number of check-ins in the area up to time $t'$. With feature $\mathbf{Fsq\_hist}$ we simply count the total number of check-ins at an area in the dataset. We refer to this group of features as $\mathbf{Fsq}$. 
\paragraph{Prediction Results:}
We perform predictions on balanced testing sets sampling an equal number of $1$s and $-1$s from different areas at each time window $t'$. 
The goal is to test if a classifier is able to discriminate from areas and times that will feature a call versus those that will not each step $t'$.
To assess the information value of a given group of features we perform two tests. In the first case, we remove it from the training set and observe the changes in the classification performance. In the second, we use only the specific group of features and no other. 

In Figure~\ref{fig:modelperform} (left panel), we note the $AUC$, $Precision$ and $Recall$ scores for the two tests. It can be observed that there are several trade-offs involved in the classifier's performance, as reflected on the precision and recall metrics. Feature group $\mathbf{Calls}$ offers the highest information value, being a set of features that is inherently more tied to the output variable than the rest. The features designed using the Foursquare data source, $\mathbf{Fsq}$, appear to add useful information to the task as not only the $AUC$ scores drop in our experiments from $0.64$ to $0.62$ as noted in Figure~\ref{fig:modelperform} (left panel). A similar trend holds for the precision and recall scores when we drop the Foursquare feature class. As noted in Figure~\ref{fig:modelperform} (right panel) where performance is examined by considering only a specific class of features,
the classifier using $\mathbf{Fsq}$ as input attains considerably higher score compared to the $\mathbf{Demo}$ (demographic) variables across all metrics. Considering the case of the $\mathbf{Demo}$ feature class, its predictive performance appears to be a balancing one, contributing towards sensitivity for the $-1$ class (no call) as classifier precision scores are consistently dependent on the presence or not of this feature class. In operational terms, both precision and recall metrics are meaningful in terms of resource utilisation. Higher recall scores imply a positive utilisation of dedicated resources to a specific spatio-temporal instance, though lower precision scores could mean that  ambulance vehicles could be planned at instances where there are no calls. We have seen that different groups of features score differently across these metrics, realising different aspects of this trade-off. 

\begin{figure}
\includegraphics[width=\linewidth]{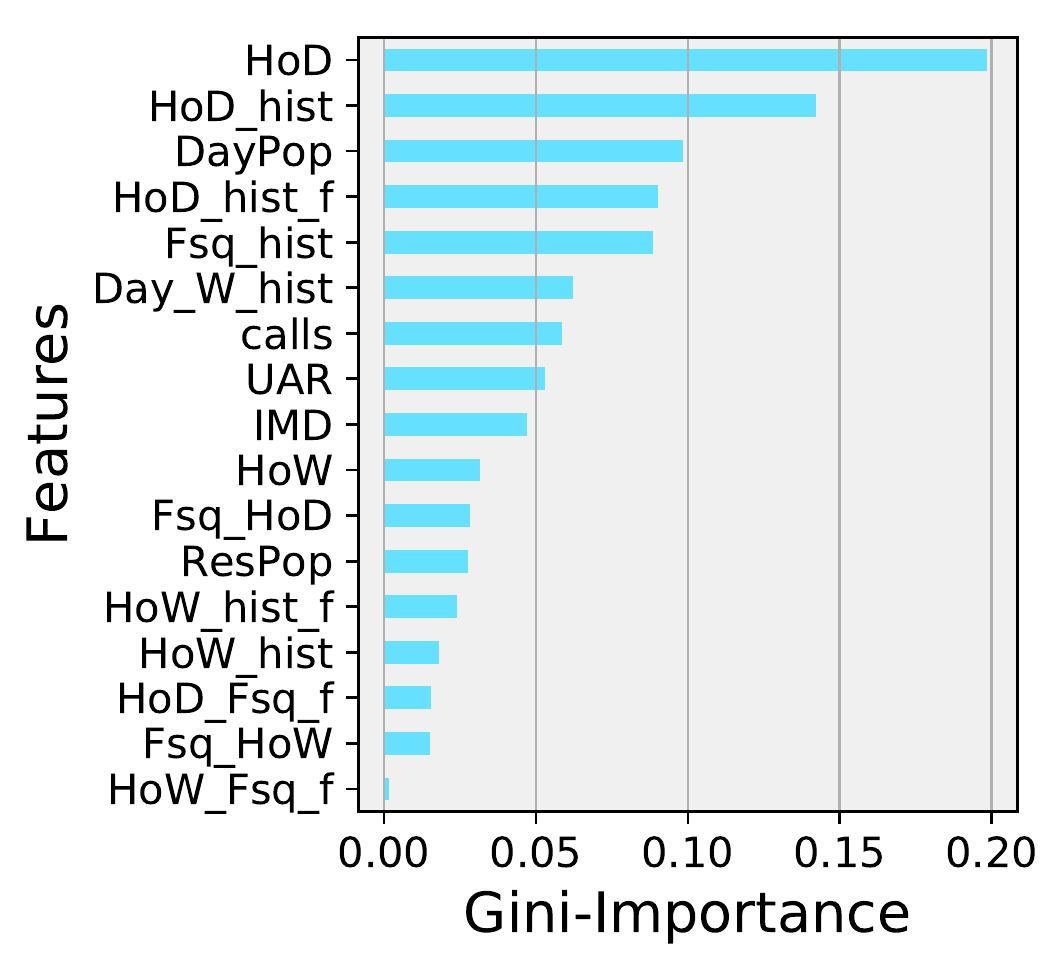}
\caption{Feature importance in terms of Gini Impurity reductions for the Random Forests model.} 
\label{fig:importance}
\end{figure}

Furthermore, in Figure~\ref{fig:importance} we plot the importance of each feature in terms of each contribution on Gini impurity reduction in the Random Forests model. The  $\mathbf{HoD}$ feature that simply notes the hour of the day offers the highest discriminative power during tree generation in the Random Forest models suggesting that the temporal patterns observed in Section~\ref{sec:analysis} are essential to predicting ambulance calls. So does the importance of the $\mathbf{HoD\_hist}$ feature suggests which counts the historic frequency of calls during the hour of the day that the model is tested.  The urban activity risk metric, which is a static metric of area characterisation, is more important than any of the $\mathbf{IMD}$ and $\mathbf{ResPop}$ features, suggesting how data from location-based services not only can shed new light in interpreting the underlying health phenomena in terms of population behavioural patterns as we noted in Section~\ref{sec:urbanrisk}, but it also becomes a promising source of data next to census data collected through government surveys. 

\begin{figure}
\includegraphics[width=\linewidth]{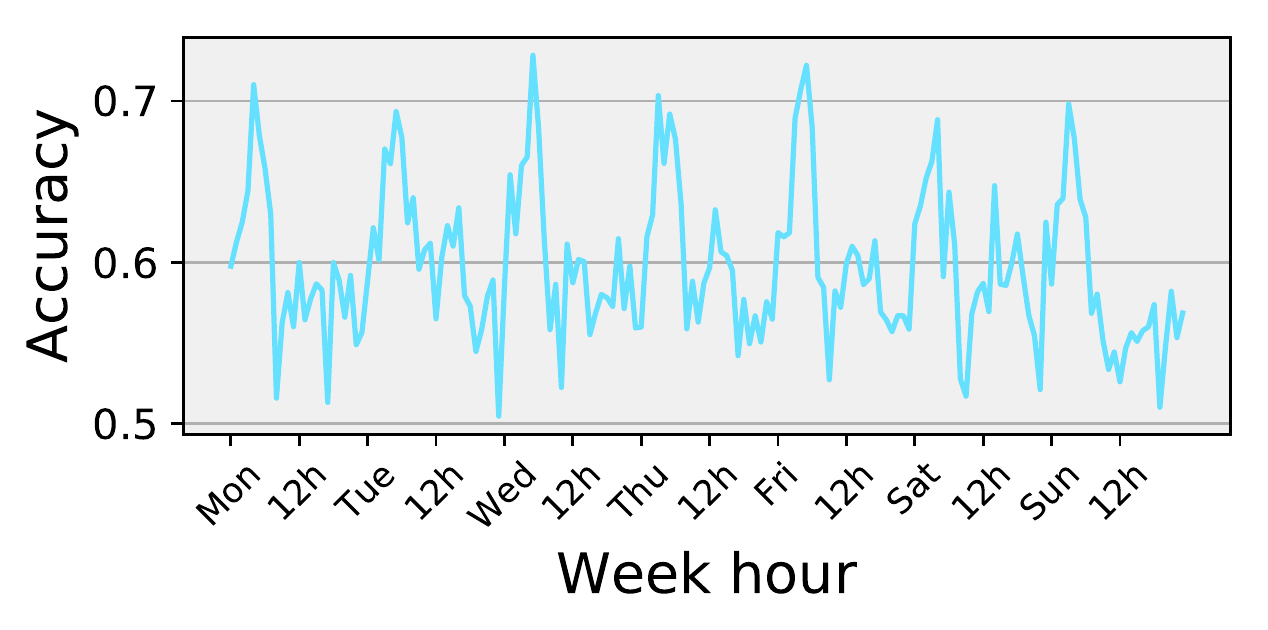}
\caption{Accuracy scores obtained by the Random Forest classifier during the course of a week.} 
\label{fig:modelweekly}
\end{figure}

Finally, given that the prediction tasks naturally unfolds through time, we performed the following further analysis. We assessed prediction performance over different hours of the week. In Figure~\ref{fig:modelweekly} we show classifier accuracy over time, defined as the fraction of correct positive or negative responses at each hour. A common pattern observed is that accuracy is higher during night hours. As also noted in Section~\ref{sec:analysis}, call frequencies drop during nighttime as people rest. The model is able to pick up this temporal pattern assuming a higher likelihood of a negative event during this period. We also note a tendency for decrease in performance during the weekend period. Populations are more likely to engage with new activities during that time and so the prediction task might become inherently harder during this time. 

%% file: conclusion.tex
\section{conclusion}
\label{sec:conclusion}
In this paper we have studied the temporal evolution of ambulance calls and corresponding incident types across a large geographic region of England. We have observed characteristic temporal patterns in health incidents and have highlighted the connection between incident prevalence at a region and its land use characteristics. The latter view has been captured through the movement of mobile users at location intelligence service Foursquare. Our study has demonstrated that ambulance services and healthcare policy design in environmental epidemiology could see benefits from the incorporation of digital datasets that describe collective user activity at fine spatio-temporal scales in online platforms. These new generation of data can augment the value extracted from datasets collected as part of the operational activity of emergency services as well as datasets collected by government sources. An inherent limitation of our work is the use of Foursquare data collected at a time period that did not directly overlap with that of the ambulance calls dataset. In future work we are planning to work with more up to date data that we expect not only to improve the prediction performance for the corresponding class of features, but also to be able to investigate the predictive power of this information source in real time prediction scenarios where dynamic population fluctuations will be captured to predict historically unseen patterns in incidence activity.
